\documentclass[ShortAfour,times,sageh]{sagej}

\usepackage{threeparttable}
\usepackage{multirow}
\usepackage{color}
\newtheorem{theorem}{Theorem}

\usepackage[normalem]{ulem}

\usepackage{moreverb,url}
\usepackage{amsmath}
\usepackage[colorlinks,bookmarksopen,bookmarksnumbered,linkcolor=blue, citecolor=blue,urlcolor=blue]{hyperref}
\usepackage{marginnote}
\usepackage{graphicx}

\newcommand*{\Scale}[2][4]{\scalebox{#1}{$#2$}}%

\newcommand\BibTeX{{\rmfamily B\kern-.05em \textsc{i\kern-.025em b}\kern-.08em
T\kern-.1667em\lower.7ex\hbox{E}\kern-.125emX}}

\newcommand{\negone}{\text{-1}}
\newcommand{\ignore}[1]{ }

\def\volumeyear{2016}

\begin{document}


\author{Timothy NeCamp\affilnum{1,2}, Amy Kilbourne\affilnum{3,4}, and Daniel Almirall\affilnum{1,2}}

\affiliation{\\ 
\affilnum{1}Department of Statistics, University of Michigan, Ann Arbor, Michigan \\
\affilnum{2}Survey Research Center, Institute for Social Research, Ann Arbor, USA \\
\affilnum{3}Department of Psychiatry, University of Michigan Medical School, Ann Arbor, USA \\
\affilnum{4}Quality Enhancement Research Initiative, HSR\&D, US Department of Veterans Affairs \\
}

\corrauth{311 West Hall, 1085 South University \\ Department of Statistics\\  University of Michigan\\Ann Arbor, MI 48109, USA
}

\email{tnecamp@umich.edu}

\begin{abstract}
Cluster-level dynamic treatment regimens can be used to guide sequential, intervention or treatment decision-making at the cluster level in order to improve outcomes at the individual or patient-level. In a cluster-level DTR, the  intervention or treatment is potentially adapted and re-adapted over time based on changes in the cluster that could be impacted by prior intervention, including based on aggregate measures of the individuals or patients that comprise it. 
Cluster-randomized sequential multiple assignment randomized trials (SMARTs) can be used to answer multiple open questions preventing scientists from developing high-quality cluster-level DTRs.
In a cluster-randomized SMART, sequential randomizations occur at the cluster level and outcomes are at the individual level. 
This manuscript makes two contributions to the design and analysis of cluster-randomized SMARTs: 
First, a weighted least squares regression approach is proposed for comparing the mean of a patient-level outcome between the cluster-level DTRs embedded in a SMART. The regression approach facilitates the use of baseline covariates which is often critical in the analysis of cluster-level trials.
Second, sample size calculators are derived for two common cluster-randomized SMART designs for use when the primary aim is a between-DTR comparison of the mean of a continuous patient-level outcome.
The methods are motivated by the Adaptive Implementation of Effective Programs Trial, which is, to our knowledge, the first-ever cluster-randomized SMART in psychiatry. 

\end{abstract}


\begin{center}
\begin{LARGE} \textbf{ Comparing cluster-level dynamic treatment regimens using sequential, multiple assignment, randomized trials: Regression estimation and sample size considerations} \par
\end{LARGE}
\end{center}

\begin{center}
\begin{large}
Tim NeCamp \footnote{tnecamp@umich.edu}\\ \end{large}
\emph{Department of Statistics, University of Michigan}\\
\emph{Survey Research Center, Institute for Social Research}\\
\ \\
\begin{large}
Amy Kilbourne\\ \end{large}
\emph{Department of Psychiatry, University of Michigan Medical School}\\
\emph{Quality Enhancement Research Initiative, HSR\&D, US Department of Veterans Affairs}\\
\ \\
\begin{large}
Daniel Almirall\\ \end{large}
\emph{Department of Statistics, University of Michigan}\\
\emph{Survey Research Center, Institute for Social Research}\\
\ \\
\begin{large} July 14th, 2016 \end{large}

\end{center}

\begin{center}
\textbf{Abstract}
\end{center} 
\begin{small}
Cluster-level dynamic treatment regimens can be used to guide sequential, intervention or treatment decision-making at the cluster level in order to improve outcomes at the individual or patient-level. In a cluster-level DTR, the  intervention or treatment is potentially adapted and re-adapted over time based on changes in the cluster that could be impacted by prior intervention, including based on aggregate measures of the individuals or patients that comprise it. 
Cluster-randomized sequential multiple assignment randomized trials (SMARTs) can be used to answer multiple open questions preventing scientists from developing high-quality cluster-level DTRs.
In a cluster-randomized SMART, sequential randomizations occur at the cluster level and outcomes are at the individual level. 
This manuscript makes two contributions to the design and analysis of cluster-randomized SMARTs: 
First, a weighted least squares regression approach is proposed for comparing the mean of a patient-level outcome between the cluster-level DTRs embedded in a SMART. The regression approach facilitates the use of baseline covariates which is often critical in the analysis of cluster-level trials.
Second, sample size calculators are derived for two common cluster-randomized SMART designs for use when the primary aim is a between-DTR comparison of the mean of a continuous patient-level outcome.  
The methods are motivated by the Adaptive Implementation of Effective Programs Trial, which is, to our knowledge, the first-ever cluster-randomized SMART in psychiatry. 

\

 \end{small}
\newpage

\section{Introduction}

Interventions aimed at improving individual-level outcomes often occur at a cluster-level \citep{raudenbush2002hierarchical,donnerklar2010clusterrandomization,murray1998design}. 
Often, it may be necessary to use a tailored, dynamic approach to intervention in order to address cluster-level heterogeneity in the kind of intervention necessary to improve individual-level outcomes \citep{kilbourne2013cluster}. 



Cluster-level dynamic treatment regimens (DTRs), also known as adaptive interventions, can be used to guide such sequential intervention decision-making at the cluster level. In a cluster-level DTR, the cluster-level intervention is potentially adapted (or re-adapted) over time based on changes in the cluster that could be impacted by prior intervention (e.g. adapting based on aggregate measures of the individuals that comprise it). 
A cluster-level DTR may also include intervention components dynamically tailored to the individuals within clusters.

Sequential multiple assignment randomized trials (SMART) represent an important data collection tool for informing how best to construct DTRs \citep{kosorok2015adaptive,lavori2014introduction,chakraborty-moodie_DTR-Book:2013,lei_SMART:2012,murphy_SMARTsim:2005}. 
The focus of most SMARTs to date has been the  development of individual-level DTRs to improve individual-level outcomes (e.g., see \citet{SMARTexMC}).  

There has been much less focus on analytic or design issues related to cluster-randomized SMARTs for developing cluster-level DTRs. In a cluster-randomized SMART, randomizations occur at the cluster level, yet outcomes are at the level of individuals within the cluster. Using the Adaptive Implementation of Effective Programs Trial (ADEPT; \citet{kilbourne2014protocol}) as a motivating example, the focus of this paper is on primary aim analysis and sample size considerations in cluster-randomized SMARTs. ADEPT, which is currently in the field, is to our knowledge the first-ever cluster-randomized SMART. The overarching goal of ADEPT is to develop a cluster-level DTR to improve the adoption of an evidence-based practice (EBP) for mood disorders in community-based mental health clinics and thereby improve patient-level mental health outcomes.

This manuscript makes two contributions to the design and analysis of cluster-randomized SMARTs. 
First, we develop a regression approach for comparing the mean of a continuous patient-level outcome between the cluster-level DTRs embedded in a SMART.  The regression approach is an extension of the estimator in \citet{nahum_SMARTprimaryPsychMeth:2012} first introduced by \citet{orellana-rot-robins_dynamicMSM_partI:2010}.  The regression approach facilitates the use of individual- and cluster-level baseline (pre-randomization) covariates in the analysis of data from a cluster-randomized SMART. 

Second, we develop sample size formulae (for the total number of clusters) to be used when the primary aim of the cluster-randomized SMART is a comparison of the mean of a continuous patient-level outcome between two DTRs beginning with different treatments. This is a common primary aim in SMARTs; see \citet[continuous end of study outcome]{oetting-bookchapter-samplesize:2011} and \citet[survival outcome]{li2011sample}.  

The regression approach can be used with any cluster-randomized SMART with repeated cluster-level randomizations. Sample size formulae are developed for two common types of two-stage SMART designs: the one used in ADEPT, and for a more common type of SMART. 

Consistent with the proposed regression approach, which facilitates the use of baseline covariates, the sample size formulae allow scientists to incorporate the correlation between a pre-specified baseline cluster-level covariate and patient-level outcomes, which leads to a reduction in the minimum number of clusters necessary \citep{spybrook2011optimal}.  This manuscript extends the work of \citet{ghosh-chakraborty-bookchap:2015}, which develops sample size calculators for a single type of cluster-randomized SMART in a non-regression context, i.e., without covariates.


\section{Sequential, Multiple Assignment Randomized Trials with Cluster-level Randomization}

Sequential multiple assignment randomized trials (SMARTs) are multi-stage randomized trial designs used explicitly for the purpose of building high-quality dynamic treatment regimens \citep{lavori-dawson_biasedRandomize:2000,murphy_SMARTsim:2005}. 
The multiple stages at which randomizations occur correspond to critical intervention decision points. 
At each decision point, randomization is used to address a question concerning the dosage (duration, frequency or amount), intensity, type, or delivery of treatment.

Here we consider SMARTs for developing cluster-level DTRs where the unit of randomization (and re-randomization) is a cluster and the outcomes are measured at the level of the individual.


\subsection{Motivating Example: The ADEPT SMART Study}

\begin{figure}
     \begin{center}
          \includegraphics[scale=0.32]{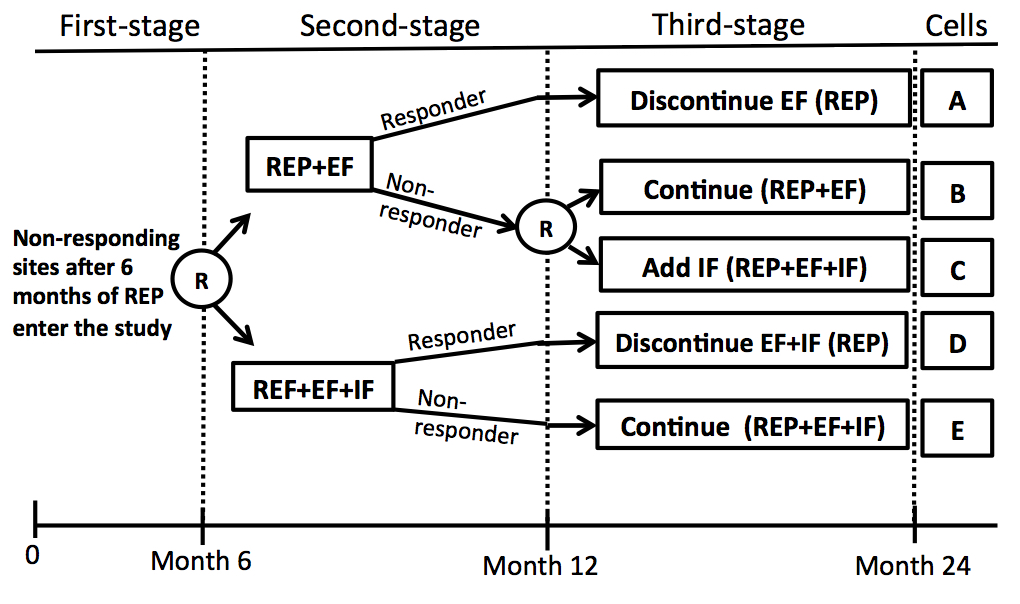}
\caption[ Schematic of ADEPT, a cluster-randomized sequential multiple-assignment randomized trial (SMART) of community-based mental health clinics.]{Schematic of ADEPT.  The encircled R signifies randomization; cluster-level randomizations occurred at baseline and after 6 months of REP+EF or REP+EF+IF following identification of clinic responder status.}\label{fig:ADEPT}
     \end{center}
\end{figure}

These methods are motivated by the Adaptive Implementation of Effective Programs Trial (ADEPT; \citet{kilbourne2014protocol} and see Figure~\ref{fig:ADEPT}), a cluster-randomized SMART in psychiatry. 
The overall aim of ADEPT is to develop a cluster-level DTR to improve the adoption of an EBP for mood disorders in community-based mental health clinics across the US states of Colorado and Michigan. The patient-level EBP is known as Life Goals \citep{kilbourne2012life}, a collaborative care, psychosocial intervention for mood disorders delivered to patients in six individual or group sessions. The primary outcome in ADEPT is a continuous, patient-level measure of mental health quality of life (MH-QOL).

ADEPT includes several interventions: the replicating effectiveness program (REP),  REP plus External Facilitation (REP+EF), and REP plus External and Internal Facilitation (REP+EF+IF).  REP is a cluster-level intervention focused on standardizing the implementation of the EBP into routine care through toolkit development, provider training, and program assistance. Facilitation is a cluster-level coaching intervention to help support the use of EBPs. External Facilitation is by phone and focuses on technical aspects of how to adopt the EBP; Internal Facilitation is in-person and involves working with a clinic manager to further embed the EBP.

ADEPT, which is currently in the field, involves community-based mental health clinics (approximately $N=60$) that have failed to respond to an initial 6 months of REP (pre-randomization).  During these 6 months, each clinic $i=1,\ldots,N$ is expected to identify approximately $m_i=10$ to $25$ patients with mood disorders, all of which are followed for patient-level outcomes throughout the study.   
Clinics that enter the study (i.e. did not respond to REP at month 6) are randomized with equal probability to receive additional REP+EF or REP+EF+IF. After another 6 months, (i) REP + EF sites that are still non-responsive are randomized with equal probability to either continue REP + EF or augment with IF (REP + EF + IF) for an additional 12 months, and (ii) facilitation interventions are
discontinued for sites that are responsive.   A clinic is identified as ``not responding" at months 6 and 12 if $<50\%$ of the patients identified to be part of Life Goals during months 0-6 have received $\ge$3 Life Goals sessions.

\begin{table}[!htbp]
\caption[The three DTRs embedded in ADEPT.]{The three DTRs embedded in ADEPT (Figure~\ref{fig:ADEPT})} 
\centering 
\begin{threeparttable}
\resizebox{\textwidth}{!}{
\begin{tabular}{c c c c c c c c c c } 
\\
\toprule

             & DTR Label                & Second-stage               & Status at end  & Third-stage& &&&Cell in& Known       \\
             & $(a_1,a_2)$              &  Treatment                & of second-stage & Treatment   & $A_1$ & R & $A_2$ & Figure & IPW \\ \hline
             & \multirow{2}{*}{$(1,1)$} & \multirow{2}{*}{REP+EF}   & Resp      & REP & 1 & 1 & & A      & 2   \\
             &                          &                           & Non Resp  & REP+EF      & 1 & 0 & 1 & B      & 4   \\ [1.ex]
             & \multirow{2}{*}{$(1,-1)$}& \multirow{2}{*}{REP+EF}   & Resp     & REP & 1 & 1 & & A      & 2   \\
             &                          &                           & Non Resp  & REP+EF+IF   & 1 & 0 & -1 & C      & 4   \\ [1.ex]
             & \multirow{2}{*}{$(-1,.)$}& \multirow{2}{*}{REP+EF+IF}& Resp      & REP & -1 & 1 & & D      & 2   \\
             &                          &                           & Non Resp  & REP+EF+IF   & -1 & 0 & & E      & 2   \\ 
\bottomrule\\
\end{tabular}}
\end{threeparttable}
\label{table:ADEPTembeddedDTRs} 
\end{table}

By design, ADEPT has three DTRs embedded within it (see Table~\ref{table:ADEPTembeddedDTRs}); each DTR is  labeled $(a_1,a_2)$.  DTR $(1,-1)$, for example, offers REP+EF at month 6; then, for clinics that remain non-responsive at month 12, REP+EF is augmented with IF; whereas, EF is discontinued for clinics who are responsive at month 12.

\subsection{The Prototypical SMART Design}

In ADEPT, only clinics not responding to REP+EF were re-randomized at the next stage. This type of SMART (but with individual-level randomizations) has been previously employed in autism research, see \citet{kasari-almirall_CCNIAautism-orangejournal:2014} and \citet{almirall2016longitudinal}.  

Many other types of SMART designs are possible (see \citet{SMARTexMC} for a comprehensive list with individual-level randomizations), including SMARTs where all units are subsequently re-randomized to the same set of next-stage intervention options (e.g., \citet{chronis2016personalized}) and others where all units are re-randomized, but to different next-stage interventions options depending on response/non-response to first-stage intervention (e.g., \citet{lu2015comparing}). Ultimately, the decision to choose a particular  type of SMART is  driven by scientific considerations. 

\begin{figure}[ht]
     \begin{center}
          \includegraphics[scale=0.32]{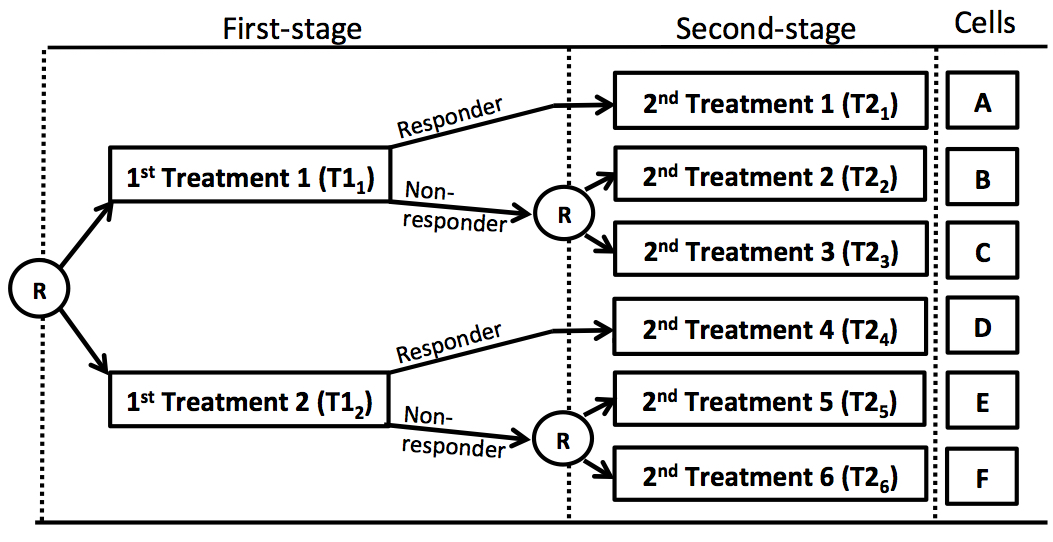}
\caption[ Schematic of a prototypical SMART design.]{Schematic of a prototypical SMART design}\label{fig:prototypical}
     \end{center}
\end{figure}

By far the most common type of SMART is a two-stage design where (i) all units are randomized to two first-stage treatment options, (ii) a subset of units at the end of stage 1 (e.g., non-responders) are re-randomized to second-stage intervention options (regardless of choice of first-stage intervention), and (iii) the remaining subset of units (e.g., responders) are not re-randomized. See Figure~\ref{fig:prototypical} for a generic example. We call this a ``prototypical SMART design" given its popularity.   Note that in the case of the prototypical SMART, there are four embedded DTRs; see Table~\ref{table:protembeddedDTRs}.

\begin{table}[!htbp] 
\caption[The four DTRs embedded in a prototypical SMART.]{The four DTRs embedded in a prototypical SMART (Figure~\ref{fig:prototypical})} 
\centering 
\begin{threeparttable}
\resizebox{\textwidth}{!}{
\begin{tabular}{c c c c c c c c c c } 
\\
\toprule

             & DTR Label                & First-stage               & Status at end  & Second-stage& &&&Cell in& Known       \\
             & $(a_1,a_2)$              &  Treatment                & of first-stage & Treatment   & $A_1$ & R & $A_2$ & Figure & IPW \\ \hline
             & \multirow{2}{*}{$(1,1)$} & \multirow{2}{*}{T$1_1$}   & Resp      & T$2_1$ & 1 & 1 & & A      & 2   \\
             &                          &                           & Non Resp  & T$2_2$      & 1 & 0 & 1 & B      & 4   \\ [1.ex]
             
             & \multirow{2}{*}{$(1,-1)$}& \multirow{2}{*}{T$1_1$}   & Resp     & T$2_1$  & 1 & 1 & & A      & 2   \\
             &                          &                           & Non Resp  & T$2_3$    & 1 & 0 & -1 & C      & 4   \\ [1.ex]
             
             & \multirow{2}{*}{$(-1,1)$}& \multirow{2}{*}{T$1_2$}& Resp      & T$2_4$  & -1 & 1 & & D      & 2   \\
             &                          &                           & Non Resp  & T$2_5$    & -1 & 0 & 1 & E      & 4   \\ [1.ex]
             
             & \multirow{2}{*}{$(-1,-1)$}& \multirow{2}{*}{T$1_2$}& Resp      & T$2_4$  & -1 & 1 & & D      & 2   \\
             &                          &                           & Non Resp  & T$2_6$   & -1 & 0 & -1 & F      & 4   \\ 
\bottomrule\\
\end{tabular}}
    \begin{tablenotes}
        \item
    \end{tablenotes}
\end{threeparttable}
\label{table:protembeddedDTRs} 
\end{table}

Published examples of the prototypical SMART (with individual-level randomizations) include \citet{pelham2016treatment} in attention-deficit/hyperactivity disorder, \citet{gunlicks2015pilot} in adolescent depression, \citet{august2014being} in conduct disorder prevention, \citet{sherwood2016bestfit} and \citet{naar2015sequential} in weight loss, and \citet{mckay2015effectofchoice} in cocaine/alcohol use.


\subsection{Common Primary Aims in a SMART}

This manuscript develops methods for comparing the mean of a continuous individual-level outcome between the DTRs embedded in a cluster-randomized SMART.  This comparison can be conceptualized in various ways as a primary aim \citep{oetting-bookchapter-samplesize:2011, almirall_IntroToSMART:2014}:
(i) To compare first stage intervention options (averaging over the second stage intervention). 
In ADEPT, this is a comparison of DTR (-1,.) vs the DTRs \{(1,1), (1,-1)\} (this was the primary aim in ADEPT; see {\citet{kilbourne2014protocol}). (ii) To compare second stage intervention options (averaging over the first stage intervention).  In the prototypical design, if the second stage treatments are the same for both first stage treatments, this would be a comparison of DTRs \{(1,1), (-1,1)\} vs DTRs \{(1,-1), (-1,-1)\} (e.g. see aim 3 in \citet{pelham2016treatment}).
(iii) To compare the mean outcome between two DTRs beginning with the same first-stage treatment. In ADEPT, this is a comparison of DTR (1,1) vs (1, -1). 
(iv) To compare the mean outcome between two DTRs that begin with different first stage treatments. In ADEPT, this is a comparison of (1,1) vs (-1,.) or of (1,-1) vs (-1,.). 

The next section develops a regression estimator that can be used to address all of these primary aims using data from a cluster-randomized SMART. Following that, we derive sample size formulae for aim (iv). Simple extensions of standard sample size formulae may be used for primary aims (i), (ii), and (iii).

\section{Methodology} 


\subsection{Marginal Mean Model}

For each SMART participant $j=1,\ldots,m_i$ within each site $i=1,\ldots,N$ we envision a primary end-of-study individual-level outcome $Y_{ij}$. Let the $p \times 1$ vector $X_{ij}$ denote a pre-specified set of baseline covariates measured prior to the initial randomization.  $X_{ij}$ may include both patient-level covariates  (e.g. age) or cluster-level covariates (e.g. clinic location).

Denote $E_{a_1,a_2}(Y_{ij}|X_{ij})$ as the marginal mean of $Y_{ij}$ had the entire population been assigned to the DTR $(a_1, a_2)$, conditional on baseline covariates, $X_{ij}$ \citep{rubin:78,neyman:1935}. $E_{a_1,a_2}(Y_{ij}|X_{ij})$ is  \emph{marginal} in that it averages over the response/non-response measure used in the DTR $(a_1,a_2)$.

Let $\mu(X_{ij},a_1,a_2;\beta, \eta)$ denote a marginal structural model \citep{orellana-rot-robins_dynamicMSM_partI:2010,murphy/laan/robins/cpprg:01,robherbrumb:00,hernan/msm-zidovudine:2000,robins:99} for the mean $E_{a_1,a_2}(Y_{ij}|X_{ij})$ which is linear in the unknown parameters $(\beta,\eta)$.  We provide examples below. $\beta$ is a $q \times 1$ vector used to denote the causal effects between the DTRs,  and $\eta$ is the $p \times 1$ vector used to denote the associational effects between $X_{ij}$ and $Y_{ij}$.

\subsubsection{Example 1: ADEPT.}  An example marginal mean model for the ADEPT study is

\begin{equation}
\mu(X_{ij}, a_1,a_2;\beta, \eta) = \beta_0 + \beta_1 a_1 + \beta_2 a_2I_{a_1=1} + \eta^TX_{ij}
\label{eq:meanADEPT}
\end{equation}
Here we use a $\beta$ with $q=3$ to capture the causal effects for the 3 embedded DTRs. $X_{ij}$ could include, for example, the three baseline site-level variables used to stratify the initial randomization: US state (Colorado vs Michigan), whether the site was a primary care or mental health site, and a site-average of individual MH-QOL scores. Using this model to address a primary aim of type (iv) above, the difference in the mean outcome had all clusters received DTR (1,1) vs had all clusters received DTR (-1,.)---i.e., $E_{1,1}(Y_{ij}) - E_{\negone, .}(Y_{ij})$---is given by $ 2\beta_1 + \beta_2$. 


\subsubsection{Example 2: Prototypical SMART.} In the prototypical SMART, we use a $\beta$ with $q=4$ to capture the causal effects for the 4 embedded DTRs.

\begin{equation}
\label{eq:meanprot}
\mu(X_{ij}, a_1,a_2;\beta, \eta)  = \beta_0 + \beta_1 a_1 + \beta_2 a_2 + \beta_3 a_1 a_2 + \eta^TX_{ij}
\end{equation}
Here, the comparison of mean outcomes had all clusters received DTR (1,1) vs had all clusters received DTR (-1,-1)---i.e.  $E_{1,1}(Y_{ij}) - E_{\negone, \negone}(Y_{ij})$---is given by $2(\beta_1 + \beta_2)$.


\subsection{Estimation}

We now present an estimator for the unknown $(\beta,\eta)$.

\subsubsection{Notation.}

Let $\textbf{X}_i$ denote the $m_i \times p$ matrix $(X_{i1}, X_{i2}, \dots, X_{im_i})^T$ of covariates and let $\boldsymbol\mu (\textbf{X}_i, a_1, a_2; \beta, \eta)$ be the $m_i \times 1$ vector of means $\left(\mu(X_{i1}, a_1, a_2; \beta, \eta), \dots, \mu(X_{im_i}, a_1, a_2; \beta, \eta) \right)^T$.  
Let $\textbf{Y}_i$ be the $m_i \times 1$ vector of responses $(Y_{i1}, Y_{i2}, \dots, Y_{im_i})^T$.  Let $A_{1i}$ denote the observed (i.e., randomly assigned) stage 1 treatment.  In ADEPT, $A_{1i} =1$ implies that cluster $i$ received REP+EF as an initial treatment while $A_{1i} = -1$ implies cluster $i$ received REP+EF+IF.   Let $R_i$, a binary variable, denote responder/non-responder status at the end of stage 1. In ADEPT, $R_i=1$ if cluster $i$ is a responder at the end of the first stage and $R_i=0$ if cluster $i$ is a non-responder.  
Let $A_{2i}$ denote the observed (i.e., randomly assigned) stage 2 treatment. Note that, depending on the SMART design, $A_{2i}$ may not be defined for some clusters $i$ depending on the value of $(A_{1i},R_i)$. In ADEPT, $A_{2i}$ is defined only for clusters with $A_{1i}=1$ and $R_i=0$. In the prototypical SMART, $A_{2i}$ is not defined for clusters with $R_{i}=1$. See Tables~\ref{table:ADEPTembeddedDTRs} and \ref{table:protembeddedDTRs}.

\subsubsection{Estimator.} Building on \citet{orellana-rot-robins_dynamicMSM_partI:2010}, \citet{nahum_SMARTprimaryPsychMeth:2012} and \citet{lu2015comparing}, we obtain $\hat\beta, \hat\eta$ by solving for $\beta, \eta$ in the following estimating equations:

\begin{multline}
0 = \sum_{i=1}^{N}U_i(A_{1i}, R_i, A_{2i}, \textbf{X}_i, \textbf{Y}_i; \beta, \eta) \triangleq \sum_{i=1}^{N}\sum_{(a_1,a_2)}  I(A_{1i}, R_i, A_{2i}, a_1, a_2) \\
\cdot W_i D(\textbf{X}_i, a_1, a_2)^TV(a_1, a_2,m_i)^{\negone}(\textbf{Y}_i - \boldsymbol\mu(\textbf{X}_i, a_1, a_2; \beta, \eta)).
\label{eq:est_eq}
\end{multline}

$D(\textbf{X}_i, a_1, a_2)$ is the $m_i \times (q+p)$ derivative of $\boldsymbol\mu(\textbf{X}_i, a_1, a_2; \beta, \eta)$ with respect to $(\beta,\eta)$; it can be thought of as the ``design matrix'' for DTR $(a_1,a_2)$. For example, using the model in Equation~\ref{eq:meanADEPT} for  ADEPT, the $j$th row of $D(\textbf{X}_i, a_1, a_2)$  is $(1, a_1, a_2I_{a_1=1}, X_{ij})$.  

$V(a_1, a_2, m_i)$ is a working model for $\text{Cov}_{a_1,a_2}(\textbf{Y}_i | \textbf{X}_i)$, the $m_i \times m_i$ covariance matrix for $\textbf{Y}_i$ conditional on $\textbf{X}_i$ for DTR $(a_1, a_2)$.  
In practice, $V(a_1, a_2, m_i)$ is unknown and must be estimated prior to solving Equation~\ref{eq:est_eq}; see Implementation section.     Note that $V(a_1, a_2, m_i)$  depends on $m_i$ through its size, not its structure.

$I(A_{1i}, R_i, A_{2i}, a_1, a_2)$ (abbreviated $I_{i (a_1,a_2)}$)  is a cluster-level indicator function which identifies whether (equals 1) or not (equals 0) cluster $i$ was assigned to a sequence of treatments that is consistent with DTR $(a_1,a_2)$.  For example, in ADEPT, if $A_{1i} = 1, R_i = 0$, and $A_{2i} = \negone$, then cluster $i$ is consistent only with DTR (1,-1); whereas if $A_{1i} = 1, R_i = 1$, then cluster $i$ is consistent with both DTR (1,1) and (1,-1). 

$W_i$ are the known cluster-level inverse probability weights (IPW, \citet{orellana-rot-robins_dynamicMSM_partI:2010}),  $W_i = 1/[Pr(A_{1i}) Pr(A_{2i}| A_{1i}, R_i)]$.  See Tables~\ref{table:ADEPTembeddedDTRs} and \ref{table:protembeddedDTRs} for the known values of $W_i$ in ADEPT and the prototypical SMART.

Following \citet{orellana-rot-robins_dynamicMSM_partI:2010}, the estimators $(\hat\beta, \hat\eta)$, derived from solving Equation \ref{eq:est_eq} are consistent and asymptotically normally distributed if the mean model (e.g., Equation \ref{eq:meanADEPT} for ADEPT) is correctly specified.  As in the generalized estimating equations literature \citep{liang1993regression,liang1986longitudinal}, there is no requirement that $V(a_1, a_2, m_i)$ be a correct model for $\text{Cov}_{a_1,a_2}(\textbf{Y}_i | \textbf{X}_i)$. See supplementary material for a sketch of the derivations.

\subsubsection{Intuition for the Weights.} By design, in the observed data in a SMART, different clusters have different probabilities of being consistent with a specific DTR. For example, clusters assigned to cells A and B are consistent with DTR (1,1)  (see Figure \ref{fig:ADEPT} and Table~\ref{table:ADEPTembeddedDTRs}). However, clusters assigned to cell A had a 50\% chance of being consistent with DTR (1,1), whereas clusters assigned to cell B had 25\% chance of being consistent with DTR (1,1). 
Ignoring this known imbalance---i.e., using an unweighted average of observations in cells A and B to estimate the mean outcome had the entire population of clusters been assigned to DTR (1,1)---would cause the Cell A observations to have an unfairly larger influence on your estimate, leading to bias. 
The weights are designed to counteract this known imbalance and ensure all clusters consistent with DTR $(a_1,a_2)$ are represented equally. For example, in ADEPT, clusters in cell A are weighted by 2, whereas clusters in cell B are weighted by 4.

\subsection{Implementation}
\label{sec:soft}

Typically, in clustered settings, our working model for $\text{Cov}_{a_1,a_2}(\textbf{Y}_i | \textbf{X}_i)$, $V(a_1, a_2, m_i)$, is taken to be ``exchangeable" and independent of $\textbf{X}_i$, i.e., $V(a_1, a_2, m_i) = \sigma^{2*}_{a_1, a_2} \cdot Exch_{m_i}(\rho^*_{a_1,a_2})$. Here $\sigma^{2*}_{a_1,a_2}$ and $\rho^*_{a_1, a_2}$ are scalars representing the conditional (on $\textbf{X}_i$) variance and intra-cluster correlation (ICC) of the outcome under DTR ($a_1,a_2$), and $Exch_{m_i}(\rho^*_{a_1,a_2}$) is an $m_i$x$m_i$ exchangeable matrix (i.e. $[Exch(\rho)]_{ii} = 1$ and $[Exch(\rho)]_{ij} = \rho$ for $i \neq j$).
The estimators $(\hat\beta, \hat\eta)$ are obtained using the following steps:

\textbf{Step 1:} Solve Equation~\ref{eq:est_eq} with $V(a_1, a_2, m_i)$ set to the identity matrix to obtain $(\hat\beta_0, \hat\eta_0$). For each embedded DTR $(a_1, a_2)$ obtain the residuals $\hat\epsilon_{ij, (a_1,a_2)}(\hat\beta_0, \hat\eta_0) = Y_{ij} - \hat\mu({X}_{ij}, a_1, a_2; \hat\beta_0, \hat\eta_0)$. 

\textbf{Step 2:} Estimate $\sigma^{2*}_{a_1,a_2}$ and $\rho^*_{a_1,a_2}$ using:
\begin{equation}
\label{eq:cor_est}
\Scale[.92]{ \hat\sigma^{2*}_{a_1,a_2} = \frac{\sum\limits_{i = 1}^N  [W_ i I_{i(a_1,a_2)} \sum\limits_{j = 1}^{m_i} \hat \epsilon^2_{ij, (a_1,a_2)} ] }{ \sum\limits_{i = 1}^N W_ i I_{i(a_1,a_2)} m_i} \,\, \text{and} \,\, \hat\rho^*_{a_1,a_2} =  \frac {\sum\limits_{i = 1}^N [W_ i I_{i(a_1,a_2)} \sum\limits_{j = 1}^{m_i} \sum\limits_{k\neq j}^{m_i} \hat \epsilon_{ij, (a_1,a_2)}  \hat \epsilon_{ik, (a_1,a_2)}]} {\hat\sigma^{2*}_{a_1,a_2} \sum\limits_{i = 1}^N W_ i I_{i(a_1,a_2)} m_i(m_i-1)}.}
\end{equation}

\textbf{Step 3: } Solve Equation~\ref{eq:est_eq} with $V(a_1, a_2, m_i)$ set to $\hat{V}(a_1, a_2, m_i) = \hat\sigma^{2*}_{a_1, a_2} \cdot Exch_{m_i}(\hat\rho^*_{a_1,a_2})$ to obtain $(\hat\beta_1, \hat\eta_1)$.

\textbf{Step 4: } Repeat Steps 2 and 3 with $\hat\epsilon_{ij, (a_1,a_2)}(\hat\beta_1, \hat\eta_1)$ to obtain final estimates $(\hat\beta, \hat\eta)$.

 In simulations we do not find appreciable performance gains by iterating Steps 2 and 3 more than twice.  Equations~\ref{eq:cor_est} can be seen as extensions of  standard working correlation estimators used in GEE literature \citep{liang1993regression,liang1986longitudinal}. 

Some analysts may choose to specify a working correlation structure which is equal for all DTR's. In this case, one could take a simple average of the estimates in Equation \ref{eq:cor_est} across all regimens $(a_1,a_2)$. 

Lastly, it is well known that by replacing the known $W_i$ in each step above with estimated weights, statistical efficiency of the estimators may be improved \citep{robins1995analysis,hernan-sim-repeatedmeasures-msm:2002,hirano2003efficient,brumback2009note,williamson2014variance,bembom:2007}

\subsection{Standard Error Estimation}  

To estimate the variance of $(\hat\beta, \hat\eta)$ we use the plug-in estimator, given by the $(q+p) \times (q+p)$ matrix $1/N \cdot \hat\Sigma_{\hat\beta, \hat\eta} = 1/N \cdot \hat J^{\negone}\hat A\hat J^{\negone}$ where

\begin{equation*}
\Scale[.95]{
\hat J = \frac{1}{N}\sum\limits_{i=1}^N\sum\limits_{(a_1,a_2)} I(A_{1i}, R_i, A_{2i}, a_1, a_2) W_i D(\textbf{X}_i,a_1,a_2)^T \hat{V}(a_1, a_2, m_i)^{\negone}D(\textbf{X}_i,a_1,a_2)}
\end{equation*}
\begin{equation*}
\begin{aligned}
\Scale[.95]{
 \hat A=\frac{1}{N}\sum\limits_{i=1}^N U_i (A_{1i}, R_i, A_{2i}, \textbf{X}_i, \textbf{Y}_i; \hat\beta, \hat\eta)U_i^T (A_{1i}, R_i, A_{2i}, \textbf{X}_i, \textbf{Y}_i; \hat\beta, \hat\eta).}
\end{aligned}
\end{equation*}
See supplementary materials for an adjustment to the standard errors for the case when weights are estimated.

\vspace{.2in}
\subsection{Hypothesis Testing} 
For any linear combination of $(\beta,\eta)$, say $c^T(\beta, \eta)$ where $c$ is a ($q+p$)-dimensional column vector, we use the univariate Wald statistic $Z= \sqrt{N} c^T(\hat\beta, \hat\eta)/ \sqrt{c^T \,  \hat{\Sigma}_{\hat{\beta}, \hat \eta}  c}$ to test the null hypothesis $H_0:c^T(\beta, \eta)=0$. For example, in ADEPT, to test the difference in means had the entire population of clusters followed DTR (1,1) versus DTR (-1,.) (i.e. primary aim (iv) above) using the model in Equation~\ref{eq:meanADEPT}, we set  $c=(0,2,1,0_p)^T$. In large samples $Z$ has a standard normal distribution under the null hypothesis. Hence, an $\alpha$ level test is ``reject $H_0$ when $|Z| > z_{\alpha/2}$," where $z_{\alpha/2}$ is the upper $\alpha/2$ quantile of a standard normal distribution.

\section{Sample Size Formulae}

For both ADEPT and the prototypical SMART, we develop sample size formulae for the total number of clusters $N$ for comparing the mean patient-level outcome between two embedded DTRs beginning with different stage 1 treatments. Specifically, for ADEPT, formulae are developed for testing null hypotheses of the form $H_0: E_{1,b_2}(Y_{ij}) - E_{\negone,.}(Y_{ij}) =0$ for a fixed $b_2 \in (-1,1)$ against alternate hypotheses of the form $H_1:E_{1,b_2}(Y_{ij}) - E_{\negone,.}(Y_{ij})=\delta \sqrt{(\sigma^2_{1, b_2} + \sigma^2_{\negone,.})/2}$. Here, $\delta$ is a standardized effect size \citep{cohen:1988} and $\sigma^2_{b_1, b_2}$ is the outcome's unconditional variance under DTR $(b_1, b_2)$.  For the prototypical SMART, formulae are developed for testing null hypotheses of the form $H_0:E_{1,b_2}(Y_{ij}) - E_{\negone,c_2}(Y_{ij})=0$ for a fixed $(b_2,c_2) \in (-1,1)^2$ against alternate hypotheses of the form $H_1:E_{1,b_2}(Y_{ij}) - E_{\negone,c_2}(Y_{ij})=\delta \sqrt{(\sigma^2_{1, b_2} + \sigma^2_{\negone,c_2})/2}$.  The formulae  are based on using (\ref{eq:est_eq}) to estimate $\beta$ in marginal models of the form (\ref{eq:meanADEPT}) or (\ref{eq:meanprot}) as follows: (i) with or without a pre-specified cluster-level covariate $X_i$, (ii) known weights $W$, and (iii) an exchangeable working covariance structure for $V$. In addition, formulae are based on a constant cluster size $m_i=m$ for all $i$ (extensions to the unequal cluster size case can be done as in \citet{kerry2001} or by conservatively setting m equal to the minimum cluster size), large sample approximations, and rely on the following \emph{working population assumptions}:

\begin{enumerate}
\item \textbf{Equal exchangeable covariance matrices across regimens:}  We  assume the true unconditional covariance matrices are equal for the two DTRs we are testing (e.g.  $ \text{Cov}_{1, b_2}(\textbf{Y}_i ) = \text{Cov}_{-1, c_2}(\textbf{Y}_i )= \sigma^2*Exch(\rho)$ in the prototypical SMART)


\item \textbf{Conditional covariance inequality:}  For a specific DTR, we assume non-responders do not vary from the marginal mean significantly more than responders. This assumption applies to different DTRs based on design, see below.   A concern about this assumption should be raised only if the scientist, apriori, believed that, for a specific DTR, non-responders had significantly larger variances than responders or if the response rate was expected to be much larger than .5 (which is atypical for SMART designs). See Appendix A for details.

\item \textbf{Correct marginal mean model:}  We assume that $E_{a_1,a_2}(Y_{ij} \mid X_{i})=\mu(X_i,a_1,a_2;\beta,\eta)$ for the pre-specified cluster-level $X_i$, where $\mu(X_i,a_1,a_2;\beta,\eta)$ is of the form (\ref{eq:meanADEPT}) or (\ref{eq:meanprot}). When $X_i$ is not included in (\ref{eq:meanADEPT}) or (\ref{eq:meanprot}), this assumption is met trivially.
\end{enumerate}

Each formula is a function of the cluster size $m$, the effect size $\delta$, the outcome's ICC, $\rho$, the probability of a cluster responding after receiving initial treatment 1, $p_1$  (i.e. $p_1=\mathbb{P}(R=1 | A_1=1)$), the probability of a cluster responding after receiving initial treatment -1, $p_{\negone}$, and the standard normal quantiles $z_{\alpha/2}$ and $z_\beta$, where $\alpha$ is the size of our test and $1-\beta$ is the power.  We first provide formulae for estimation without covariates followed by the case when a cluster-level covariate $X_i$ is used.

\subsection{ADEPT Sample Size Formula}

For ADEPT, working assumption 2 is: $E_{1,b_2}[(\textbf{Y}_i-\boldsymbol\mu{(1, b_2)})(\textbf{Y}_i-\boldsymbol\mu{(1, b_2)})^T|R=0] \preceq E_{1,b_2}[(\textbf{Y}_i-\boldsymbol\mu{(1, b_2)})(\textbf{Y}_i-\boldsymbol\mu{(1, b_2)})^T] = \text{Cov}_{1, b_2}(\textbf{Y}_i)$.   Also, for ADEPT, working assumption 1 can be relaxed to $\sigma^2_{1,b_2} \leq \sigma^2_{-1,.}$. Under these assumptions we obtain the sample size formula:

\begin{equation}
\label{eq:samp_size_adept}
N = \frac{4(z_\beta + z_{\alpha/2})^2}{m\delta^2} \cdot (1+(m-1)\rho) \cdot (1+\frac{1-p_1}{2})  
\end{equation}

\subsection{Prototypical Sample Size Formula}

For Prototypical SMART designs, working assumption 2 is: for both DTRs in our test, i.e. $(a_1, a_2) = (1, b_2) \text{ and } (-1, c_2)$,   $E_{a_1, a_2}[(\textbf{Y}_i-\boldsymbol\mu{(a_1, a_2)})(\textbf{Y}_i-\boldsymbol\mu{(a_1, a_2)})^T|R=0] \preceq E_{a_1, a_2}[(\textbf{Y}_i-\boldsymbol\mu{(a_1, a_2)})(\textbf{Y}_i-\boldsymbol\mu{(a_1, a_2)})^T] = \text{Cov}_{a_1, a_2}(\textbf{Y}_i)$.  
Under these assumptions we obtain the sample size formula:

\begin{equation}
\label{eq:samp_size_prot}
N = \frac{4(z_\beta + z_{\alpha/2})^2}{m\delta^2} \cdot (1+(m-1)\rho)  \cdot (1 + \frac{(1-p_1)+(1-p_\negone)}{2}) 
\end{equation}

\noindent
Note this formula is identical to the formula in \citet{ghosh-chakraborty-bookchap:2015}.

The sample size formulae in Equations \ref{eq:samp_size_adept} and \ref{eq:samp_size_prot} are intuitive.  The first two terms in both formulae are identical; these terms comprise the formulae for the sample size for the difference in means in a 2-arm randomized control trial (RCT) with cluster-level randomization \citep{donnerklar2010clusterrandomization}.  The second term, in particular, is the expression for the variance inflation factor (VIF) arising from cluster-randomized trials. If $\rho$ = 0 (i.e. VIF = 1), there is no inflation due to cluster randomization because we have no correlation within clusters. As $\rho$ increases, each new observation within a cluster provides less unique information causing the VIF to increase. This, in turn, leads to an increase in sample size, $N$. 

The third term, which is unique to SMARTs, is used to account for the fact that some clusters are re-randomized depending on response at the end of stage 1; hence, this last term is a function of the rate of response to first stage intervention.  To understand this third term, it is useful to consider two extremes in the context of the prototypical SMART: If both response rates ($p_1, p_{\negone}$) are 1, then there is no re-randomization and the design is analogous to a 2-arm cluster-randomized RCT (here, the third term is equal to 1).  If, on the other hand, both response rates are 0, then all clusters are randomized twice; here, the third term is equal to 2. Note how the the third term is different for ADEPT and the prototypical SMART due to the difference in randomization schemes.  Also, in the special case where response rates to initial treatments are equal (i.e. $p_1 = p_{\negone}$), we would end up with the clustered version of the sample size formula in \citet{oetting-bookchapter-samplesize:2011}.

\subsection{Sample Size Formula with a Cluster-level Covariate}

When including a cluster-level covariate in (\ref{eq:meanADEPT}) or (\ref{eq:meanprot}), working assumption 2 is similar for each corresponding design, except it involves the conditional (on $\textbf{X}_i$) marginal mean, i.e. $E_{a_1, a_2}[(\textbf{Y}_i-\boldsymbol\mu{(\textbf{X}_i, a_1, a_2}))(\textbf{Y}_i-\boldsymbol\mu{(\textbf{X}_i, a_1, a_2}))^T|R=0] \preceq E_{a_1, a_2}[(\textbf{Y}_i-\boldsymbol\mu{(\textbf{X}_i, a_1, a_2}))(\textbf{Y}_i-\boldsymbol\mu{(\textbf{X}_i, a_1, a_2}))^T].$  Also, our formula depends on $\text{Cor}(Y,X)$, which is the is the scalar correlation between the outcome $Y_{ij}$ and the cluster-level covariate $X_i$ under the DTRs in our test.  Note that under assumptions 1 and 3, this correlation is constant across these DTRs. We obtain the following sample size formula for ADEPT:
\begin{equation}
\label{samp_size_cov}
N =  \frac{4(z_\beta + z_{\alpha/2})^2}{m\delta^2} \cdot (1+(m-1)\rho^*) \cdot  (1+\frac{1-p_1}{2}) \cdot [1-\text{Cor}^2(Y,X)]
\end{equation}
For the prototypical SMART, the sample size formula is:
\begin{equation}
\label{eq:samp_size_cov2}
N =  \frac{4(z_\beta + z_{\alpha/2})^2}{m\delta^2} \cdot (1+(m-1)\rho^*)  \cdot (1 + \frac{(1-p_1)+(1-p_\negone)}{2}) \cdot [1-\text{Cor}^2(Y,X)]
\end{equation}
where $\rho^* = \frac{\rho-\text{Cor}^2(Y,X)}{1-\text{Cor}^2(Y,X)}$

The use a covariate leads to two changes in the sample size formulae. First, as expected, depending on the strength of the correlation between $X$ and $Y$ (i.e., $\text{Cor}^2(Y,X)$), the use of a covariate has the potential to reduce the minimum required sample size; this is because the use of covariates may improve the efficiency of our estimator of $\beta$.  Second, there is a reduction in sample size due to the reduction in correlation, $\rho^*$, which, by definition, is always less than $\rho$.

\subsection{Using the Sample Size Formula for the ADEPT study}

To exemplify how the formula can be utilized in practice, we calculate the sample size needed to detect a difference between DTRs (1,-1) and (-1,.) in ADEPT. This difference would help us understand if it is better to give REP+EF+IF to non-responding clinics initially, or to delay REP+EF+IF until a clinic is non-responsive to REP+EF. In ADEPT, we expect the ICC of patient's MH-QOL to be $\rho = .01$ and the probability of responding when initially receiving REP+EF to be $p_1 = .2$.  Using the true sample size of $N = 60$, a common cluster size of $m = 10$, and performing an $\alpha =$ .05 level test ($z_{\alpha/2} = 1.96$), by rearranging our formula, we conclude that at 80\% power ($z_\beta = .84$) we can detect an effect size of $\delta$ = .282.



\section{Simulations}

Simulations were conducted to evaluate the developed formulae and understand their robustness to violations of the working assumptions.   
Specifically, we evaluate formulae under four scenarios: (1) satisfying all working assumptions, (2) violating working assumption 1, (3) violating working assumption 2, and (4) violating working assumption 3. 
Here, we present results for ADEPT; results were similar for the prototypical SMART.

Details concerning the data generative model can be found in Appendix B.  Data were generated to mimic the ADEPT study. We considered different data generative scenarios with varied standardized effect sizes ($\delta$ = .2 (small), .5 (moderate)), cluster sizes ($m = 5, 10, 20$), ICC ($\rho$ or $\rho^*$ = .01 or .1), and, when there is a cluster-level covariate, the correlation between $X$ and $Y$ (Cor$^2(Y, X) \in [.04, .4]$). We also considered different scenarios constituting violations of the working assumptions (details below).  For each scenario, sample size was selected based on the proposed formulae with power ($1-\beta$) = .9, $\alpha=0.05$. 1000 data sets were generated for each scenario.  

Each data set was analyzed as in the Implementation and Standard Error Estimation sub-sections, using the marginal mean model in Equation~\ref{eq:meanADEPT}.  For each scenario, we compared the estimated power (over 1000 data sets) to nominal power of .9.






\begin{table}[ht]
\caption[No violation, no covariates, ADEPT]{Power analysis of Formula \ref{eq:samp_size_adept} } 
\centering 
\begin{threeparttable}
\resizebox{\textwidth}{!}{
\begin{tabular}{c c c c c c c c}
\\
\toprule

             &ICC, $\rho$                & Effect Size, $\delta$  & Cluster Size, m  & Sample Size, N & Assumptions & Violating & Violating \\
             & &  &   &  & are correct &  Assumption 1 & Assumption 2\\
             
              \hline
             &	.01	 & .2		 & 5 	&   306	 &.894  & .891 & .886 \\
             &	&		 &  20	& 	88	 &.917  & .890 & .876* \\
             &	&	.5	 &  5	&	49	 	 &.909  & .898 & .880* \\
             &	&	 &  10 	&	26	 	 &.906  &  .878* & .893\\
             &	.1	 & .2		 & 5 	&   412	 &.910  & .901 & .870*\\
             &	&		 &  20	& 	213	 &.922*  & .902 & .891 \\
             &	&	.5	 &  5	&	66	 	 &.909  & .888 & .898\\
             &	&	 &  20 	&	34	 	 &.915  & .913 & .889 \\
\bottomrule
\end{tabular}}
    \begin{tablenotes}
        \item \footnotesize{*The proportion is significantly different from .9 at the 5\% level.}
    \end{tablenotes}
\end{threeparttable}
\label{table:sim_no_cov_no_vio} 
\end{table}

Table~\ref{table:sim_no_cov_no_vio} describes simulation results for the sample size formula in Equation~\ref{eq:samp_size_adept}. To violate assumption 1, we made the response variance under DTR$(1,b_2)$  1.5 times the response variance under DTR $(-1,.)$.  We could have also violated this assumption by deviating from an exchangeable covariance structure, however, in cluster-randomized trials it is rare to use an other covariance structure \citep{eldridge2009}. To violate assumption 2, we made non-responders have significantly larger variance than responders under DTR $(1,b_2)$.

As expected, when no assumptions are violated (column 5), our estimated power is close to our pre-specified power, .9.  When assumption 1 is violated (column 6) or assumption 2 is violated (column 7), we see that our power does not reduce dramatically.  Hence, we conclude that our sample size formula is robust to violations of working assumptions 1 and 2.  

Since working assumption 3 will always be true when there are no covariates, we run a second simulation to evaluate the robustness of the sample size formula in Equation \ref{samp_size_cov} (i.e. with a cluster-level covariate) to this assumption.  Specifically, to violate assumption 3, we deviate from the linear marginal mean in Equation~\ref{eq:meanADEPT} by generating data with $E_{a_1,a_2}(Y_{ij}|X_{i}) =  \beta_0 + \beta_1 a_1 + \beta_2 a_2I_{a_1=1} + \eta f_k(X_{i})$ where $f_k(X_{i}) = X_{i} \text{ for } X_{i} \in [-k,k],\  f_k(X_{i}) = k \text{ for } X_{i} > k, \text{ and } f_k(X_{i}) = - k \text{ for } X_{i} < - k$ (i.e. the linear marginal mean is misspecified outside of $[-k,k]$).  Here $\eta$ is chosen to maintain the same values of $\text{Cor}(Y,X)$.  Setting $k = 2$ indicates a small violation (column 7) and setting $k = 1$ indicates a large violation (column 8).  We still, however, analyze the data using the marginal mean model in Equation~\ref{eq:meanADEPT}.  The results are in Table~\ref{table:sim_cov}.  

\begin{table}[ht]
\caption[Covariates, ADEPT]{Power analysis of Formula \ref{samp_size_cov} } 
\centering 
\begin{threeparttable}
\resizebox{\textwidth}{!}{
\begin{tabular}{c c c c  c c c c c}
\\
\toprule

             &ICC, $\rho^*$                & Effect Size, $\delta$  & Cluster Size, m  & Cor$^2(Y,X)$ & Sample Size, N & Assumptions & Small & Large \\ 
             && &  &   &  & are correct &  Violation of & Violation of \\ 
             && &  &   &  &  &  Assumption 3 & Assumption 3 \\ \hline
             &	.01	 & .2		 & 5 	& .238 &    233	 & .909 & .904 & .859* \\
             &	&		 &  20	& .238  & 	65	 & .903 & .879* & .777* \\
             &	&	.5	 &  5	&	.043	 	 &  47 & .891 & .901 & .902 \\
             &	&	 &  10 	&	.066	 	 & 24 &$.893$ & .903 & .897 \\
             &	.1	 & .2		 & 5 	&    .243	 &  305 & .918  & .900 &.890  \\
             &	&		 &  20	&  .243	& 159 & .915  & .922* & .864* \\
             &	&	.5	 &  5	&		.043 	 & 63 & .898 & .916 & .919* \\
             &	&	 &  20 	&		.043 	 &  32 &   .908 & .920* & .899\\
\bottomrule
\end{tabular}}
    \begin{tablenotes}
        \item \footnotesize{*The proportion is significantly different from .9 at the 5\% level.}
    \end{tablenotes}
\end{threeparttable}
\label{table:sim_cov} 
\end{table}

As expected, when no assumptions are violated (column 6), our estimated power is close to our pre-specified power, .9. Note the reduction in sample size caused by the addition of a covariate. Under a small violation (column 7), we see the power is not significantly reduced.  Under a large violation (column 8), we see our power is lowest when X and Y are moderately correlated and the sample size is low. This is because when X and Y are weakly correlated, the overall influence of X is small, and hence misspecification of the relationship between X and Y will have little influence on our estimation and power.


\section{Discussion and Future Work}

This manuscript presents a regression estimator and sample size formulae for comparing embedded dynamic treatment regimens using data arising from a cluster-randomized SMART.  Methods were motivated by the ADEPT SMART, a study designed to develop a dynamic treatment regimen (at the level of community based mental health clinics) designed to improve mental health outcomes for patients clustered within those sites \citep{kilbourne2014protocol}. Sample size formulae were derived for both ADEPT and for a more common type of SMART. 


There are a number of directions for future research in the analysis of cluster-randomized SMARTs.  
First, relatively staightforward applications of the estimator in Equation \ref{eq:est_eq} with different link functions can be used to analyze, for example, binary, count or zero-inflated outcomes.

Second, in practice, many cluster-randomized SMARTs will collect longitudinal (i.e., repeated measures) research outcomes at the patient-level. A natural next step is to combine the estimator presented here with methods for the analysis of longitudinal SMART outcomes \citep{lu2015comparing} in order to accomodate two levels of clustering: repeated measures within patients within clusters. 

Third, future work could also consider the use of variance components models, i.e., mixed effects or random effects models \citep{hedeker2006longitudinal,raudenbush2002hierarchical}, which are now-standard in the analysis of randomized trials.

Fourth, while this manuscript focuses on the analysis of primary aims in a SMART, in the DTR literature there is much interest in the development and application of analysis methods designed to generate hypotheses about more individually-tailored DTRs \citep{zhao2015new,laber2014sizing,zhang2015usingdecisionlists,laber2015tree,linn2014interactive,qian2011performance,moodie2014qlearningflexible,zhou2015residual}.  Much of this literature has focused on identifying optimal DTRs at the individual level. Such methods could be extended for the analysis of data arising from cluster-randomized SMARTs to develop optimal cluster-level DTRs.

There are also a number of interesting methodological issues related to the design of cluster-randomized SMARTs (with implications for analysis methods). First, the sample size formulae derived here were limited to cases where our data contains a single cluster-level covariate. Future work may provide extensions to data containing multiple covariates and individual-level covariates. 

Second, in this manuscript we focus on SMARTs that are useful for developing of cluster-level DTRs where the initial and subsequent decisions are all at the cluster-level. 
However, there is currently much interest by educational scientists in SMARTs aimed at developing DTRs where sequences of intervention decisions are made at \emph{both} the cluster and individual level. 
For example, we are currently involved in the conduct of a trial where the first stage intervention is at the level of classrooms with children with autism (such classrooms often include 1 to 3 children with autism), and the subsequent stages of intervention are at the level of the children themselves \citep{kasariIESgrantPilotSMART}. 


\begin{acks}
R code for implementing the weighted least squares regression estimator and the sample size formulae for both ADEPT and the prototypical SMART are provided on the first author's website. This research is supported by the
		following NIH grants:
		R01MH099898 (Kilbourne \& Almirall),
		P50DA039838 (Almirall), 
		R01HD073975 (Almirall),
		R01DA039901 (Almirall).
We also would like to thank Xi (Lucy) Lu for the helpful comments.
\end{acks}


\bibliography{refs}
\bibliographystyle{SageH.bst}


\section{Appendix A: Derivation of Sample Size Formulae}\label{app:samplesize}

Here we derive the sample size formula.  We begin with deriving the sample size formula for a prototypical design with no covariates. Then we make simple extensions to formula under an ADEPT design and/or when a cluster-level covariate is included.  All sample size formulae are based primary aim (iv), that is the marginal mean comparison of two dynamic treatment regimens that begin with a different initial treatment (i.e., $E_{1,b_2}(Y_{ij}) - E_{\negone,c_2}(Y_{ij})$).

Also, for simplicity, these derivations assume the marginal mean model is parameterized as follows:

For Prototypical:
\begin{equation*}
 \mu(X_{ij}, a_1,a_2;\beta, \eta)  = \beta_0I\{a_1 = 1, a_2 = 1\} + \beta_1I\{a_1 = 1, a_2 = -1\} + 
 $$ $$
 \beta_2I\{a_1 = -1, a_2 = 1\} + \beta_3I\{a_1 = -1, a_2 = -1\} + \eta^TX_{ij}
 \end{equation*}

For ADEPT:
\begin{equation*} 
\mu(X_{ij}, a_1,a_2) = \beta_0I\{a_1 = 1, a_2 = 1\} + \beta_1I\{a_1 = 1, a_2 = -1\} + 
$$ $$
\beta_2I\{a_1 = -1\} + \eta^TX_{ij}
\end{equation*}

Fitting the re-parameterized models will yield the exact same conclusions as fitting the marginal mean models in Equations \ref{eq:meanADEPT} and \ref{eq:meanprot}.

\subsection{Prototypical Design without Covariates}

For data arising from a prototypical SMART design, we derive the sample size formula for detecting a significant difference between mean outcomes from two treatment regimens, $(b_1, b_2)$ and $(c_1, c_2)$. Since we are  interested in comparing two regimes starting with different initial treatment.  Hence, without loss of generality, assume $b_1 = 1$ and $c_1 = \negone$.

We are interested in the following hypothesis test:

\begin{center}
$H_0: E_{1,b_2}(Y_{ij}) - E_{\negone,c_2}(Y_{ij}) = 0 $
\end{center}
against the alternative
\begin{center}
$H_1: E_{1,b_2}(Y_{ij}) - E_{\negone,c_2}(Y_{ij}) =  \delta\sigma$
\end{center}
where $\sigma = \sqrt{\frac{\sigma_{1, b_2}^2 + \sigma_{\negone, c_2}^2}{2}}$, $ \sigma_{a_1, a_2}^2 =\text{Var}_{a_1,a_2}(Y_{ij})$, and $\delta$ is the standardized effect size.

We make a series of assumptions to derive our sample size formulae.  We highlight these assumptions throughout our derivation to illustrate there use in the calculation.  Note that our sample size is developed for a fixed cluster size, $m$ (extensions to the unequal cluster size case can be done as in \citet{kerry2001}).

Our test statistic used for the hypothesis test is 
\begin{equation*}
Z =  \frac{\sqrt{N}(\hat\mu{(1, b_2)} - \hat\mu{(\negone, c_2)})}{\sqrt{\hat\tau^2{(1, b_2)  + \hat\tau^2{(\negone, c_2) } - 2\widehat{Cov}( \sqrt{N}\hat\mu{(1, b_2)}, \sqrt{N}\hat\mu{(\negone, c_2)}) }} } 
\end{equation*}

Here, $\hat\tau^2{(a_1, a_2)}$ is an approximation of the variance, $\tau^2{(a_1, a_2)}$ = Var($\sqrt{N} \hat\mu{(a_1, a_2)}$) as given in the supplementary material.

In large samples and under assumption 3 described in the Sample Size Formulae section, the distributions of $\hat\mu{(1, b_2)}$ and $\hat\mu{(\negone, c_2)}$ can be approximated by the normal distribution, $\hat\tau^2{(1, b_2)} \approx \tau^2{(1, b_2)}$, $\hat\tau^2{(\negone, c_2)} \approx \tau^2{(\negone, c_2)}$, and $ \widehat{Cov}( \sqrt{N}\hat\mu{(1, b_2)}, \sqrt{N}\hat\mu{(\negone, c_2)}) \approx  {Cov}( \sqrt{N}\hat\mu{(1, b_2)}, \sqrt{N}\hat\mu{(\negone, c_2)}) = 0$ (here the covariance is 0 due to the independence of estimators of marginal means with different initial treatments).  Thus, $Z$ approximately has a standard normal distribution under the null hypothesis.  

Note that these calculations are exactly the same as those highlighted in the Hypothesis Testing section.  Specifically, under the original parameterization, letting $c = (0, 2, b_2 - c_2, b_2 + c_2,0_p)^T$ then $c^T(\hat\beta, \hat\eta) = \hat\mu{(1, b_2)} - \hat\mu{(\negone, c_2)}$ and $c \hat{\Sigma}_{\hat{\beta}, \hat \eta} \, c^T = \hat\tau^2{(1, b_2)} + \hat\tau^2{(\negone, c_2)} -2\widehat{Cov}( \sqrt{N}\hat\mu{(1, b_2)}, \sqrt{N}\hat\mu{(\negone, c_2)})$.

Under the alternative, our test statistic is normal with approximate mean $ \sqrt{N}\delta\sigma/\ \sqrt{\tau^2{(1, b_2)} + \tau^2{(\negone, c_2)}} $ and variance 1.  Doing standard power calculations \citep{oetting-bookchapter-samplesize:2011} for a hypothesis test of size $\alpha$, in order to obtain desired power of $1-\beta$, we need to find $N$ that satisfies:
\begin{equation*}
z_\beta \approx -z_{\alpha/2} + \frac{\delta\sigma\sqrt{N}}{\sqrt{\tau^2{(1, b_2)} + \tau^2{(\negone, c_2)}}}
\end{equation*}
\begin{equation}
N = \frac{(z_\beta + z_{\alpha/2})^2(\tau^2{(1, b_2)} + \tau^2{(\negone, c_2)})}{\delta^2\sigma^2}
\label{eq:N}
\end{equation}

Everything in this formula can be explicitly found except $\tau^2{(1, b_2)}$ and  $\tau^2{(\negone, c_2)}$.  Hence we now aim to derive upper bounds for these variables in order to write our sample size formula in terms of either known or easily elicited quantities.

Note that under the parameterization, for any DTR $(a_1, a_2)$, $\tau^2{(a_1, a_2)} = \text{Var}[\sqrt{N}\hat\mu{(a_1, a_2)}] = [\Sigma_{\hat \beta}]_{(a_1,a_2)} = [J^{\negone} E[U_i U_i^T]J^{\negone}]_{(a_1,a_2)} $ , with $U_i$ and $J$ defined in the supplementary material.  Here, for ease, for the 4x4 matrices, M = $\Sigma_{\hat \beta}$, J, or E[$U_iU_i^T$], we define $[M]_{(a_1, a_2)}$ as the diagonal element corresponding to DTR $(a_1, a_2)$  (e.g. the (3,3) element for DTR $(-1,1)$). Also, $1_m$ is defined as the $m$x1 vector of 1's. Lastly, as defined in the Sample Size Formulae section, $p_{a_1} $ is the probability of responding given the cluster had received initial treatment $a_1$.

After simplification, we find J is a diagonal matrix with diagonal element:
\begin{equation*}
[J]_{(a_1, a_2)} = E[W_i I_{i (a_1, a_2 )}1_m^TV^\negone( a_1, a_2)1_m] = 1_m^TV^\negone( a_1, a_2)1_m
\end{equation*}

 For $E[U_iU_i^T]_{(a_1, a_2)}$, we perform the following simplification:
 \begin{equation*}
 E[U_iU_i^T]_{(a_1, a_2)} = \\
 $$ $$
  E[W_i^2 I_{i (a_1, a_2 )} 1_m^TV^\negone(  a_1, a_2)(\textbf{Y}_i-\boldsymbol\mu{(a_1, a_2)})(\textbf{Y}_i-\boldsymbol\mu{(a_1, a_2)})^TV^\negone(  a_1, a_2)1_m]  =\\
  $$ $$
 1_m^TV^\negone(  a_1, a_2)E_{a_1,a_2}[W_i(\textbf{Y}_i-\boldsymbol\mu{(a_1, a_2)})(\textbf{Y}_i-\boldsymbol\mu{(a_1, a_2)})^T]V^\negone(  a_1, a_2)1_m  =\\
  $$ $$
 1_m^TV^\negone(  a_1, a_2)[2E_{a_1,a_2}[(\textbf{Y}_i-\boldsymbol\mu{(a_1, a_2)})(\textbf{Y}_i-\boldsymbol\mu{(a_1, a_2)})^T|R=1]p_{a_1} + \\ 
 $$ $$ 
 4E_{a_1,a_2}[(\textbf{Y}_i-\boldsymbol\mu{(a_1, a_2)})(\textbf{Y}_i-\boldsymbol\mu{(a_1, a_2)})^T|R=0](1-p_{a_1})]V^\negone(  a_1, a_2)1_m = \\
 $$ $$
2*1_m^TV^\negone(  a_1, a_2)\Sigma_{a_1, a_2}V^\negone(  a_1, a_2)1_m +\\
$$ $$ 
2(1-p_{a_1})*1_m^TV^\negone(  a_1, a_2)E_{a_1,a_2}[(\textbf{Y}_i-\boldsymbol\mu{(a_1, a_2)})(\textbf{Y}_i-\boldsymbol\mu{(a_1, a_2)})^T|R=0]
$$ $$
\ \cdot V^\negone(  a_1, a_2)1_m 
\end{equation*}

To go from line 2 to 3, we assume Robin's consistency assumption holds, i.e. that the cluster's observed outcomes equal the cluster's potential outcomes under the observed DTR \citep{robins:97}. Under this assumption we are able to switch from $E$, which is an expected value over observed data, to $E_{a_1,a_2}$ which is the expected value had the entire population received DTR ($a_1,a_2$) \citep{rubin:78,neyman:1935}.  

For further simplification, we now make assumption 2. This assumption is equivalent to assuming, for a specific DTR ($a_1, a_2$) (we drop the $a_1, a_2$ from the subscripts for convenience):
$|(\sigma^2_R\rho_R - \sigma^2_{NR}\rho_{NR})p_{a_1} + (\mu_R - \mu_{NR})^2(p_{a_1})(1-2p_{a_1})| \leq (\sigma^2_R - \sigma^2_{NR})p_{a_1} + (\mu_R - \mu_{NR})^2(p_{a_1})(1-2p_{a_1})$, where $\mu_R, \sigma^2_R, \rho_R$ are the mean, variance, and ICC of responders had the whole population received DTR ($a_1, a_2$), (i.e. $\mu_R = E_{a_1,a_2}(Y_{ij} | R_i = 1),  \sigma^2_R =  \text{Var}_{a_1,a_2}(Y_{ij} | R_i = 1), \rho_R = \text{Cov}_{a_1,a_2}(Y_{i1}, Y_{i2} | R_i = 1)/\text{Var}_{a_1,a_2}(Y_{ij} | R_i = 1) $), similarly defined for NR and non-responders (i.e. conditional on R = 0).  Also, $p_{a_1}$ is the probability of response, given initial treatment $a_1$.

For DTR ($a_1, a_2$), this condition is satisfied if the probability of response is less than or equal to .5 (which is typical for prototypical SMART designs), the non-responders of that regimen have a variance which is less than or equal to the variance of responders of the regimen, and both responders and non-responders have similar within cluster covariances.  


Under assumption 2 we can bound and simplify our expression for  $ E[U_iU_i^T]_{(a_1, a_2)} $ as
\begin{equation*}
E[U_iU_i^T]_{(a_1,a_2)} \preceq \\
$$ $$
2(1 + (1-p_{a_1}))*1_m^TV^\negone(a_1, a_2)\Sigma_{a_1, a_2}V^\negone(a_1, a_2)1_m 
\end{equation*}

We next utilize the fact that our working covariance matrix, V, is exchangeable.  With some linear algebra, this assumption allows us to perform the following simplification: $ \frac{2(2-p_{a_1})1_m^TV^\negone(  a_1, a_2)\Sigma_{a_1, a_2}V^\negone(  a_1, a_2)1_m}{(1_m^TV^\negone(  a_1, a_2)1_m)^2} = \frac{2(2-p_{a_1})1_m^T\Sigma_{a_1, a_2}1_m}{m^2}$

Next, using assumption 1, we exploit the exchangeable population covariance structure (i.e. $\text{Cov}_{a_1, a_2}(\textbf{Y}_i )= \sigma_{a_1,a_2}^2*Exch(\rho_{a_1,a_2}$), where $\rho_{a_1,a_2} = \text
{Cor}_{a_1,a_2}(Y_{i1}, Y_{i2})$). Putting everything together, we obtain:
\begin{equation*}
\tau^2{(a_1, a_2)} =  \text{Var}[\sqrt{N}\hat\mu{(a_1, a_2)}] \stackrel{3}{=} [J^{\negone} E[U_iU_i^T]J^{\negone}]_{(a_1, a_2)} \stackrel{2}{\leq} 
$$ $$
\frac{2(2-p_{a_1})1_m^TV^\negone(  a_1, a_2)\Sigma_{a_1, a_2}V^\negone(  a_1, a_2)1_m}{(1_m^TV^\negone(  a_1, a_2)1_m)^2} \stackrel{}{=} \frac{2(2-p_{a_1})1_m^T\Sigma_{a_1, a_2}1_m}{m^2} \stackrel{1}{=}
\end{equation*}
\begin{equation} \label{tauequation}
 \frac{2(2-p_{a_1})\sigma^2_{a_1, a_2}[1+(m-1)\rho_{a_1, a_2}]}{m}
\end{equation}

We utilize the across regimen covariance equality of assumption 1 in order to simplify things further. Note that if one had good estimates of $\sigma^2_{a_1, a_2} \text{ and } \rho_{a_1, a_2}$, then you could easily obtain $N$ by plugging in the values into equation \ref{tauequation} and then using these estimates in equation \ref{eq:N}.

Using this equality, we combine equation \ref{tauequation} with equation \ref{eq:N} and simplify to obtain:
\begin{equation*}
N = \frac{4(z_\beta + z_{\alpha/2})^2}{m\delta^2} \cdot (1+(m-1)\rho)  \cdot (1 + \frac{(1-p_1)+(1-p_\negone)}{2}) 
\end{equation*}

\subsection{ADEPT Design without Covariates}

All the calculations done above are nearly identical for the ADEPT case.   The only major difference arises from the lack of re-randomization of clusters receiving initial treatment $a_1 = -1$.  This in fact makes the calculations simpler for $\tau^2(-1,.)$  In particular, we assume assumptions 1 and 3, however, assumption 2 only needs to be assumed for DTR (1,$b_2$).  Under these assumptions we obtain:

\begin{equation}
\tau^2{(1, b_2)} \leq \frac{2(2-p_{1})\sigma^2_{1, b_2}[1+(m-1)\rho_{1, b_2}]}{m}, 
\tau^2{(\negone,.)} = \frac{2\sigma^2_{\negone,.}[1+(m-1)\rho_{\negone,.}]}{m} 
\label{eq_tau}
\end{equation}

After utilizing the across regimen population covariance equality of assumption 1, we combine the equation \ref{eq_tau} with equation \ref{eq:N} and simplify to obtain:
\begin{equation*}
N = \frac{4(z_\beta + z_{\alpha/2})^2}{m\delta^2} \cdot (1+(m-1)\rho) \cdot (1+\frac{1-p_1}{2})   
\end{equation*}

\noindent
For this sample size formula, we actually only need to assume $\sigma^2_{1,b_2} \leq \sigma^2_{-1,.}$ (as opposed to the equality assumed in assumption 1) to ensure our power is larger than 1 - $\beta$.

\subsection{With a Cluster-Level Covariate}

In this section, we write the sample size formula when adding a single cluster-level covariate to the model (i.e. the $m$x1 vector $\textbf{X}_i \triangleq (X_{i1}, X_{i2}, \dots, X_{im})^T = (X_{i}, X_{i}, \dots, X_{i})^T)$.   First, for DTR ($a_1,a_2)$, we define  $\sigma^{2*}_{a_1,a_2} \triangleq E_{a_1,a_2}[\text{Var}_{a_1,a_2}(Y_{ij}|X_i)] $, and $\rho^*_{a_1,a_2} \triangleq E_{a_1,a_2}[\text{Cov}_{a_1,a_2}(Y_{i1},Y_{i2}|X_i)]/\sigma^{2*}_{a_1,a_2} $, where the expectations are taken over $X_i$.  Note in the homoscedastic case, the expectation is unnecessary since the conditional variances and covariances are constant for all $X_i$. 

The key to extending our formulae to the covariate case is observing that the numerator of our sample size formula will now be in terms of the average \emph{conditional} variances and ICCs, $\sigma^{2*}_{a_1,a_2} $ and $\rho^*_{a_1,a_2}$, while the denominator remains in terms of the overall variance, $\sigma^{2}_{a_1,a_2}$.  Since $\sigma^{2}_{a_1,a_2} =  \sigma^{2*}_{a_1,a_2} + \eta^2Var(X_{i})$ and $\text{Cov}_{a_1,a_2}(Y_{i1},Y_{i2}) = E_{a_1,a_2}[\text{Cov}_{a_1,a_2}(Y_{i1},Y_{i2}|X_i)] +  \eta^2Var(X_{i})$, 
then  $\sigma^{2*}_{a_1,a_2} $ and $\rho^*_{a_1,a_2}$ must be less than or equal to $ \sigma^{2}_{a_1,a_2}$ and $\rho_{a_1,a_2}$.  Thus the numerator of our formula is reduced while the denominator remains the same.  With some algebra, this reduction is shown to be $1-\text{Cor}^2(Y_{ij},X_i)$. Note that this reduction can be shown to be the same reduction arising from including a cluster-level covariate in clustered RCTs, see \citet{hedges:09}.

For simplification, we do all calculations assuming our covariate has mean 0 (this eliminates covariance between our marginal mean estimates).  When our covariate does not have mean 0, one can show that mean centering our covariate does not change the value of our test statistic, and hence does not change our power.  Thus our sample size formula remains valid when the covariate does not have mean 0.

\subsubsection{Prototypical Design}

Using assumptions 1-3, the relationships above, and doing similar algebra as in the non-covariate case (except now everything is in terms of $\sigma^{2*}_{a_1,a_2}$ and $\rho^*_{a_1,a_2}$), we obtain for both DTRs of interest:

\begin{equation}
\label{eq:prot_cov}
\tau^2{(a_1, a_2)} \leq \frac{2(2-p_{1})\sigma^{2*}_{a_1, a_2}[1+(m-1)\rho^*_{a_1, a_2}]}{m}
\end{equation}

Making assumption 1 (on unconditional population variances and correlations) will lead to equality of expected conditional variances and correlations due to the simple relationship between the conditional and unconditional variances and covariances highlighted above.  Hence we define $ \rho^*  \triangleq \rho^*_{1, b_2} = \rho^*_{-1, c_2} $  and  $ \sigma^{2*}  \triangleq \sigma^{2*}_{1, b_2} = \sigma^{2*}_{\negone,c_2} $.

We also take advantage of the fact that $ \text{Cor}^2_{1,b_2}(Y_{ij},X_i) = \eta^2\text{Var}(X_i)/\sigma^2_{1,b_2} = \eta^2\text{Var}(X_i)/\sigma^2_{-1,c_2}$, i.e. is also equal across both regimes. We define $\text{Cor}^2(Y,X) \triangleq \text{Cor}^2_{1,b_2}(Y_{ij},X_i) = \text{Cor}^2_{-1,c_2}(Y_{ij},X_i)$.

Ultimately, we obtain :

\begin{equation*}
N =  \frac{4(z_\beta + z_{\alpha/2})^2}{m\delta^2}(1+(m-1)\rho^*)(1 + \frac{(1-p_1)+(1-p_\negone)}{2})[1-\text{Cor}^2(Y,X)]
\end{equation*}

\subsubsection{ADEPT Design}

Similar to the prototypical design, under assumptions 1-3, for DTR (1, $b_2$) we obtain the same bound as in equation \ref{eq:prot_cov}.  For DTR (-1, .), without utilizing assumption 2, we obtain:

\begin{equation*}
\tau^2{(-1,.)} = \frac{2\sigma^{2*}_{-1,.}[1+(m-1)\rho^*_{-1,.}]}{m} 
\end{equation*}

And, utilizing equality of population covariance across regimens:
\begin{equation*}
N =  \frac{4(z_\beta + z_{\alpha/2})^2}{m\delta^2}(1+(m-1)\rho^*)(1+\frac{1-p_1}{2})[1-\text{Cor}^2(Y,X)]   
\end{equation*}

Also, with some algebra, $\rho^*$ can be expressed as $\rho^* = (\rho-\text{Cor}^2(Y,X))/(1-\text{Cor}^2(Y,X))$, allowing our two covariate sample size formulae to be a function purely of $\rho$ and Cor$^2$(Y,X).

%
%
%

\section{Appendix B: Data-generative Models Used in Simulation Experiments}
Below we describe how we generated data for our simulations.

\subsection{Without Covariates}

For Table~\ref{table:sim_no_cov_no_vio}, we generate data, ($A_1, R, A_2, Y$), for each of the $N$ clusters as follows:
\begin{enumerate}
\item Generate $A_1$ to be 1 or -1 with equal probability

\item Generate $R$ to be 1 with probability $p_{A_1}$ and 0 otherwise

\item Generate $A_2$ to be 1 or -1 with equal probability, for clusters with $A_1 = 1$, $R = 0$

\item Generate the $m$x$1$ vector $Y= \mu_{A_1, R, A_2} + \epsilon$, where $\epsilon \sim$ MVN($0, \Sigma_{A_1, R, A_2})$, where $ \Sigma_{A_1, R, A_2} = \sigma^2_{A_1, R, A_2} \cdot Exch_m(\rho_{A_1, R, A_2}).$ Here $\mu_{A_1, R, A_2},\   \sigma^2_{A_1, R, A_2},\  \rho_{A_1, R, A_2}$ are the cell means, variances, and ICCs since they correspond to each cell in Figure \ref{fig:ADEPT}.

\end{enumerate}

\begin{table}[ht]

\caption[Vio5 no covariates, ADEPT]{Pre-specified simulation values for Table~\ref{table:sim_no_cov_no_vio}} 
\centering 
\begin{threeparttable}
\resizebox{\textwidth}{!}{
\begin{tabular}{c c c c c c c c c c}
\\
\toprule

Simulation &$p_1$ & $p_{\negone}$ & $\mu_{1, 1, .}$  & $ \mu_{1, 0, 1}$ & $\mu_{1, 0, -1}$ & $\mu_{-1, 1, .}$ &   $\mu_{-1, 0, .}$ & $ \sigma^2_{1, 1, .} 	$ & 		\\ \hline
Table~\ref{table:sim_no_cov_no_vio}, Row 1, Col 5 & .2 & .3 & 34.71 & 32.71 & 28 & 32.7 & 31 & 63.36 & \\
Table~\ref{table:sim_no_cov_no_vio}, Row 1, Col 6 & .2& .3 & 34.71 & 32.71 &  28 &  32.14 &  31.44 & 63.36 & \\
Table~\ref{table:sim_no_cov_no_vio}, Row 1, Col 7 & .2 & .3 & 33.36 &  33.05  & 28 & 32.7 & 31 & 1 & \\
\hline
Simulation  & $ \sigma^2_{1, 0, 1}$ & $\sigma^2_{1, 0, -1}$ & $\sigma^2_{-1, 1, .}$ &   $\sigma^2_{-1, 0, .}$ & $\rho_{1, 1, .}$  & $ \rho_{1, 0, 1}$ & $\rho_{1, 0, -1}$ & $\rho_{-1, 1, .}$ &   $\rho_{-1, 0, .}$	\\
\hline
Table~\ref{table:sim_no_cov_no_vio}, Row 1, Col 5 & 63.36 & 60 & 63.39 & 63.39 & 0.0 &  0.0 &  0.0 &  .0006 &  .0006 \\
Table~\ref{table:sim_no_cov_no_vio}, Row 1, Col 6 & 63.36 & 60 & 43 & 43 & 0.0 &  0.0 &  0.0 &  .0076 &  .0076 \\
Table~\ref{table:sim_no_cov_no_vio}, Row 1, Col 7 & 79.73 & 60 & 63.39 & 63.39 & 0.9 &  .007 &  0.0 &  .0006 &  .0006 \\
\bottomrule
\end{tabular}}
\end{threeparttable}
\label{table:gen_no_cov} 
\end{table}

Under these specified means, variances, and ICCs, one can easily obtain the desired marginal (over $R$) means, variances, and ICCs under a specific DTR using the laws of total expectation and variation.  For example, to obtain the marginal mean under DTR (1,1), one would calculate $\mu_{1,1,.} p_1 + \mu_{1,0,1}  (1-p_1)$.  To calculate the variance under DTR (1,1), one would calculate $\sigma^2_{1,1,.}p_1 + \sigma^2_{1,0,1}(1-p_1) + p_1(1-p_1)(\mu_{1,1,.}-\mu_{1,0,1} )^2$. To calculate the covariance under DTR (1,1), one would calculate  $\sigma^2_{1,1,.}\rho_{1,1,.}p_1 + \sigma^2_{1,0,1}\rho_{1,0,1}(1-p_1) + p_1(1-p_1)(\mu_{1,1,.}-\mu_{1,0,1} )^2$


When no assumptions were violated (row 1 of Table \ref{table:gen_no_cov}), the cell means and variances were first chosen to give marginal means and variances which are both similar to results expected in ADEPT and produce effect sizes matching Table \ref{table:sim_no_cov_no_vio}.  After obtaining the correct effect size, the cell ICCs were then chosen also to match values specified in Table \ref{table:sim_no_cov_no_vio}.  To violate assumptions (row 2 and 3 of Table \ref{table:gen_no_cov}), the cell means, variances, and ICCs from row 1 were slightly altered to create the correct violations.

\subsection{With a Cluster-Level Covariate}

To generate data for Table~\ref{table:sim_cov}, we use a continuous cluster-level covariate.  We generate data, ($X, A_1, R, A_2, Y$), for each of the $N$ clusters as follows:

\begin{enumerate}
\item Generate $A_1$ to be 1 or -1 with equal probability

\item Generate $R$ to be 1 with probability $p_{A_1}$ and 0 otherwise

\item Generate $A_2$ to be 1 or -1 with equal probability, for clusters with $A_1 = 1$, $R = 0$

\item Generate a single cluster-level covariate $X$ from Normal(0,1)

\item 
\subitem a. Generate  $m$x$1$ vector $Y= \mu_{A_1, R, A_2} + \eta X + \epsilon$ for Column 6
\subitem b. Generate $m$x$1$ vector $Y= \mu_{A_1, R, A_2} + \eta f_k(X) + \epsilon$ for Columns 7, 8

where $\epsilon \sim$ MVN($0, \Sigma_{A_1, R, A_2})$, with $ \Sigma_{A_1, R, A_2} = \sigma^{2*}_{A_1, R, A_2} \cdot Exch_m(\rho^*_{A_1, R, A_2}).$ Here $\mu_{A_1, R, A_2},\   \sigma^{2*}_{A_1, R, A_2},\  \rho^*_{A_1, R, A_2}$ are the cell means, conditional cell variances, and conditional cell ICCs since they correspond to each cell in Figure \ref{fig:ADEPT}.  Also, $f_k$ is the same piecewise function defined in the Simulations section (i.e. which is non-linear outside of $[-k,k]$).
\end{enumerate}

\begin{table}[ht]

\caption[Vio5 no covariates, ADEPT]{Pre-specified simulation values for Table~\ref{table:sim_cov} } 
\centering 
\begin{threeparttable}
\resizebox{\textwidth}{!}{
\begin{tabular}{c c c c c c c c c c c}
\\
\toprule

Simulation & k &$p_1$ & $p_{\negone}$ & $\eta$&$\mu_{1, 1, .}$  & $ \mu_{1, 0, 1}$ & $\mu_{1, 0, -1}$ & $\mu_{-1, 1, .}$ &   $\mu_{-1, 0, .}$ & $ \sigma^{2*}_{1, 1, .} 	$ 		\\ \hline
Table~\ref{table:sim_cov}, Row 1, Col 6 &  & .2 & .3 & 4.47 & 34.94 & 32.94 & 28 & 32.7 & 31 & 63.36 \\
Table~\ref{table:sim_cov}, Row 1, Col 7 & 2 & .2 & .3 & 4.69 & 34.95 & 32.95 & 28 & 32.7 & 31 & 63.36 \\
Table~\ref{table:sim_cov}, Row 1, Col 8 & 1 & .2 & .3 & 6.66 & 34.98 & 32.98 & 28 & 32.7 & 31 & 63.36 \\
\hline
Simulation  & $ \sigma^{2*}_{1, 0, 1}$ & $\sigma^{2*}_{1, 0, -1}$ & $\sigma^{2*}_{-1, 1, .}$ &   $\sigma^{2*}_{-1, 0, .}$ & $\rho^*_{1, 1, .}$  & $ \rho^*_{1, 0, 1}$ & $\rho^*_{1, 0, -1}$ & $\rho^*_{-1, 1, .}$ &   $\rho^*_{-1, 0, .}$	&\\
\hline
Table~\ref{table:sim_cov}, Row 1, Col 6 & 63.36 & 60 & 63.39 & 63.39 & 0.0 &  0.0 &  0.0 &  .0006 &  .0006 & \\
Table~\ref{table:sim_cov}, Row 1, Col 7 & 63.36 & 60 & 63.39 & 63.39 & 0.0 &  0.0 &  0.0 &  .0006 &  .0006 & \\
Table~\ref{table:sim_cov}, Row 1, Col 8 & 63.36 & 60 & 63.39 & 63.39 & 0.0 &  0.0 &  0.0 &  .0006 &  .0006 & \\
\bottomrule
\end{tabular}}
\end{threeparttable}
\label{table:gen_cov} 
\end{table}

Under these specified conditional means, variances, and ICCs, one can again obtain the desired conditional and unconditional marginal means, variances, and ICCs under a specific DTR using the laws of total expectation and variation.  For example, to obtain the conditional marginal variance under DTR (1,1), one would calculate $\sigma_{1,1}^{2*} = \sigma_{1,1,.}^{2*}p_1 + \sigma_{1,0,1}^{2*}(1-p_1) + p_1(1-p_1)(\mu_{1,1,.}-\mu_{1,0,1} )^2$. For data generated as in 5a, to obtain the unconditional marginal variance under DTR (1,1), one would calculate $\sigma_{1,1}^{2} = \sigma_{1,1}^{2*} +  \eta^2\text{Var}(X)$.  For data generated as in 5b, we instead calculate $\sigma_{1,1}^{2} = \sigma_{1,1}^{2*} +  \eta^2\text{Var}(f_k(X))$. Both $ \text{Var}(X) $ and $\text{Var}(f_k(X))$ can be found using the known distribution of $X$. 

The cell means, conditional variances, and conditional ICCs were chosen for the same reason as in the non-covariate case. $\eta$ was chosen to give the correct correlation between $X$ and $Y$.

\newpage


\section{Supplementary Materials}
\subsection{Asymptotic results for the estimator}

 

This section shows consistency and asymptotic normality of the proposed estimator.  These proofs are similar to those found in \citet{lu:2016}.  In Equation \ref{eq:est_eq} the estimator was presented with fixed working covariance matrices, $V(a_1,a_2, m_i)$, and known weights, $W_i$. However, in practice $V(a_1,a_2, m_i)$ must be estimated and the known weights can be estimated to improve efficiency, see \citet{robins1995analysis,hernan-sim-repeatedmeasures-msm:2002,hirano2003efficient,brumback2009note,williamson2014variance,bembom:2007}.  Additionally, we may want the covariance matrices to be functions of baseline covariates.  We also may want to allow weights to depend on baseline covariates, $\textbf{X}_i$, information collected prior to first randomization, $\text{L}_{0i}$, and information collected between the first and second randomization, $\text{L}_{1i}$.   Specifically, we allow $W_i = 1/(Pr(A_{1i} | \textbf{X}_i, \text{L}_{0i})Pr(A_{2i}| \textbf{X}_i, \text{L}_{0i}, A_{1i}, \text{L}_{1i}, R_i))$. We represent $V(a_1,a_2, m_i)$ and $W_i$ by $V(\textbf{X}_i, a_1,a_2, m_i; \hat\alpha)$ and $W_i(\hat\gamma)$, where $\hat\alpha$ and $\hat\gamma$ arise from estimation of $V$ and $W$.  We also allow for cluster sizes to be unequal across observations since this is typical in practice.  Under these general settings, the estimating equation is: 
\begin{equation} \label{eq:est_eq2}
\begin{aligned}
 0 =  \frac{1}{N}\sum\limits^N_{i=1}\sum\limits_{(a_1,a_2)} & I(A_{1i}, R_i, A_{2i}, a_1, a_2) W_i( \hat\gamma) \\ & \cdot  D(\textbf{X}_i,a_1,a_2)^TV(\textbf{X}_i, a_1,a_2, m_i; \hat\alpha)^{\negone}(\textbf{Y}_i - \boldsymbol\mu(\textbf{X}_i, a_1, a_2; \beta, \eta)),
\end{aligned}
\end{equation}
We first demonstrate the consistency of the estimator found by solving this equation.  

\begin{theorem} \label{thm:consistency}
Assume the marginal model is correctly specified, that is, $E_{a_1, a_2}[Y_{ij}|X_{ij}] = \mu(X_{ij},a_1,a_2;\beta_0,\eta_0)$, where $(\beta_0, \eta_0)$ is the true value for the parameter $(\beta, \eta)$ in the marginal mean model.  Assume $Y_{ij}$ is conditionally independent of $X_{ik}$ $\forall k \neq j$ given $X_{ij}$.  Also assume that there exists $\alpha^+, \gamma_0$ such that $\sqrt{N}(\hat\alpha - \alpha^+)  = O_p(1)$ and $\sqrt{N}(\hat\gamma - \gamma_0) = O_p(1)$ (i.e. are bounded in probability), where $W_i( \gamma_0) \equiv W_i$, the true inverse-probability weight.  Then the estimator $(\hat\beta, \hat\eta)$ obtained by solving Equation~\ref{eq:est_eq2} is consistent for $(\beta_0, \eta_0)$.
\end{theorem}

Proof.  Define $\theta = (\beta,\eta)$ to denote the marginal mean model parameters with true values $\theta_0 = (\beta_0, \eta_0)$. We denote the estimating equation in \ref{eq:est_eq2} as $0 =  1/N\sum^N_{i=1}U_i(Z_i; \theta, \hat\alpha, \hat\gamma),$ where $Z_i$ is all observed covariates and responses for cluster i.  It remains to show that $E[U_i(Z_i; \theta_0, \alpha^+, \gamma_0)] = 0_{p+q}$, from which consistency can be established as done for the standard GEE estimator \citep{liang1986longitudinal}.  Here the expectation is over observed data (with respect to the distribution of the observed data, $P_{obs}$)  as opposed to $E_{a_1, a_2}$, which is an expectation over data arising as if all clusters had received DTR ($a_1,a_2$) (with respect to the distribution $P_{a_1, a_2}$).

Note that $I(A_{1i}, R_i, A_{2i}, a_1, a_2) /W_i$ is the Radon-Nikodym derivative between $P_{obs}$ and $P_{a_1,a_2}$.   And thus,
\begin{equation*} 
\begin{aligned}
&E[U_i(Z_i; \theta_0, \alpha^+, \gamma_0)]  = \\
&\sum\limits_{(a_1,a_2)}E_{a_1,a_2}[D(\textbf{X}_i, a_1,a_2)^TV(\textbf{X}_i, a_1, a_2, m; \alpha^+)^{\negone}(\textbf{Y}_i - \boldsymbol\mu(\textbf{X}_i, a_1, a_2; \theta_0))]\\ 
&\sum\limits_{(a_1,a_2)}E_{\textbf{X}_i}[D(\textbf{X}_i,a_1,a_2)^TV(\textbf{X}_i, a_1, a_2, m; \alpha^+)^{\negone}]E_{a_1,a_2}[\textbf{Y}_i - \boldsymbol\mu(\textbf{X}_i, a_1, a_2; \theta_0)|\textbf{X}_i] \\
& = 0_{p+q}
\end{aligned}
\end{equation*}

\noindent
The final equation equals zero due to the conditional independence and correct specification of the marginal mean model.  

We next prove the asymptotic normality of our estimator obtained in equation~\ref{eq:est_eq2}.  We borrow notation from the previous proof.

\begin{theorem}
Assuming mild regularity conditions, the same assumptions as in Theorem~\ref{thm:consistency}, the cluster sizes are bounded, and that the weight parameter $\gamma$ is obtained from maximum likelihood estimation for treatment assignment probabilities, with a score function $S_\gamma $.  Then $\sqrt{N}((\hat\beta, \hat\eta)-(\beta_0, \eta_0))$ is asymptotically multivariate normal with zero mean and covariance matrix $\Sigma_{\hat\beta, \hat\eta} = J^{\negone}(A - CB^\negone C^T)J^{\negone}$, where A, B, C, and J are given by
\begin{equation*}
\begin{aligned}
J = \lim_{N\to\infty}\frac{1}{N}\sum\limits_{i=1}^NE\sum\limits_{(a_1,a_2)} & I(A_{1i}, R_i, A_{2i}, a_1, a_2)  W_i(\gamma_0) \\ & \cdot D(\textbf{X}_i,a_1,a_2)^TV(\textbf{X}_i, a_1, a_2, m_i; \alpha^+)^{\negone}D(\textbf{X}_i,a_1,a_2)
\end{aligned}
\end{equation*}
\begin{equation*}
\begin{aligned}
 A=\lim_{N\to\infty}\frac{1}{N}\sum\limits_{i=1}^NE[U_iU_i^T], B=\lim_{N\to\infty}\frac{1}{N}\sum\limits_{i=1}^NE[S_{\gamma_0,i}S_{\gamma_0,i}^T],  C=\lim_{N\to\infty}\frac{1}{N}\sum\limits_{i=1}^NE[U_iS_{\gamma_0,i}^T]
\end{aligned}
\end{equation*}
with $U_i \triangleq U_i(Z_i; \theta_0, \alpha^+, \gamma_0)$
\end{theorem}
 
Proof: Using the same argument for GEE estimators \citep{liang1986longitudinal} we obtain
\begin{equation*}
\begin{aligned}
\sqrt{N}(\hat\theta-\theta_0) &= \left[\lim_{N\to\infty}\frac{1}{N}\sum\limits^N_{i=1}\frac{\partial U_i(Z_i; \theta_0, \alpha^+, \gamma_0)}{\partial \theta}\right]^{\negone}  \left\{ \frac{1}{\sqrt{N}}\sum\limits^N_{i=1}U_i(Z_i; \theta_0, \alpha^+, \gamma_0) \right. \\
 &\left. + \left[\lim_{N\to\infty}\frac{1}{N}\sum\limits^N_{i=1}\frac{\partial U_i(Z_i; \theta_0, \alpha^+, \gamma_0)}{\partial \gamma}\right]\sqrt{N}(\hat\gamma - \gamma_0) \right\} + o_p(1)
\end{aligned}
\end{equation*}

Using the fact that $S_\gamma$ is the score function for $\hat\gamma$ to express $\sqrt{N}(\hat\gamma - \gamma_0)$ as a sum.  Also, using the fact that our cluster size is bounded combined with the Law of Large Numbers, we write all long-run averages of random variables as long run averages of expectations. Hence, we obtain:

\begin{equation*}
\begin{aligned}
&\sqrt{N}(\hat\theta-\theta_0) =  \left[\lim_{N\to\infty}\frac{1}{N}\sum\limits^N_{i=1}E\left\{ \frac{\partial U_i}{\partial \theta}\right\} \right]^{\negone}
\left\{ \frac{1}{\sqrt{N}}\sum\limits^N_{i=1}U_i - \right. \\ 
& \left. \left[\lim_{N\to\infty}\frac{1}{N}\sum^N_{j=1}E(U_jS^T_{\gamma_0,j})\right]
\left[\lim_{N\to\infty}\frac{1}{N}\sum^N_{j=1}E(S_{\gamma_0,j}S^T_{\gamma_0,j})\right]^{\negone} 
S_{\gamma_0,i}\right\} + o_p(1) \\
& = J^\negone \frac{1}{\sqrt{N}}\sum\limits^N_{i=1}(U_i - CB^\negone S_{\gamma_0,i}) + o_p(1) \rightarrow Normal[0,  J^{\negone}(A - CB^\negone C^T)J^{\negone}]
\end{aligned}
\end{equation*}

Remark: Note that with unequal cluster sizes, $U_i$ and $S_{\gamma,i}$ are not identically distributed, and hence we must express our variances with long run averages.  If cluster sizes were equal, averaging would not be necessary and we would obtain $A = E[U_iU_i^T]$, $B = E[S_{\gamma_0}S_{\gamma_0}^T], C = E[U_iS_{\gamma_0}^T]$, and $J = E[\partial U_i/ \partial \theta]$.

To obtain estimates for our standard error of $(\hat\beta, \hat\eta)$ we use plug in estimates of A, B, C, and J.  Specifically, we set:
\begin{equation*}
\begin{aligned}
\hat J =\frac{1}{N}\sum\limits_{i=1}^N\sum\limits_{(a_1,a_2)}& I(A_{1i}, R_i, A_{2i}, a_1, a_2)  W_i(\hat\gamma) \\ &\cdot D(\textbf{X}_i,a_1,a_2)^TV(\textbf{X}_i, a_1, a_2, m_i; \hat\alpha)^{\negone}D(\textbf{X}_i,a_1,a_2)
\end{aligned}
\end{equation*}
\begin{equation*}
\begin{aligned}
 \hat A=\frac{1}{N}\sum\limits_{i=1}^N\hat U_i\hat U_i^T, \hat B=\frac{1}{N}\sum\limits_{i=1}^NS_{\hat\gamma,i}S_{\hat\gamma,i}^T,  \hat C=\frac{1}{N}\sum\limits_{i=1}^N\hat U_iS_{\hat\gamma,i}^T
\end{aligned}
\end{equation*}
where $\hat U_i =  U_i(Z_i; \hat\theta, \hat\alpha, \hat\gamma) $

Thus, the plug in estimator for $\Sigma_{\hat\theta}$ is $\hat\Sigma_{\hat\theta} = \hat J^{\negone}(\hat A - \hat C\hat B^\negone \hat C^T)\hat J^{\negone}$ and we obtain $\widehat{Var}(\hat\theta) = 1/N \cdot \hat\Sigma_{\hat\theta} $.

%
%
%

\end{document}